# Increasing the Detectability of Phase-Amplitude Coupling

Mojtaba Chehelcheraghi[1,2], Chie Nakatani[1], Cees van Leeuwen[1,3]

Corresponding author: mojtaba.chehelcheraghi@kuleuven.be

Authors Email: chie.nakatani@kuleuven.be, cees.vanleeuwen@kuleuven.be

1. Brain and Cognition Unit, KU Leuven, Leuven, Belgium

2. Present Address: Telecommunications and Microwaves Unit (TELEMIC), Department of Electrical Engineering (ESAT), KU Leuven, Leuven, Belgium

3. Center for Cognitive Science, TU Kaiserslautern, Kaiserslautern, Germany

**Funding:** This research is founded by an Odysseus grant from the Flemish organizations for science (FWO) awarded to Cees van Leeuwen.

**Abbreviations**

MCA: Modulatory Component Analysis

CFC: Cross Frequency Coupling

PAC: Phase Amplitude Coupling

PLV: Phase Locking Value

EPS: Envelope Phase Synchronization

MVL: Mean Vector Length

CV: Coherence Value

KLD: Kullback-Leibler Distance

AMI: Amplitude Modulation Index

LFP: Local Field Potential

PSD: Power Spectral Density

SNR: Signal to Noise Ratio

**Abstract**

**Background:** In electrical brain signals such as Local Field Potential (LFP) and Electroencephalogram (EEG), oscillations emerge as a result of neural network activity. The oscillations extend over several frequency bands. Between their dominant components, various couplings can be observed. Of these, Phase-Amplitude Coupling (PAC) is intensively studied in relation to brain function. In the time-frequency domain, however, PAC measurement faces a dilemma in the choice of filter bandwidth. For a frequency m modulating a frequency n, filters narrowly tuned around the latter frequency will miss the modulatory components at frequencies n+m and n-m; wide band tuning will pass increasing levels of noise.

**New Method:** Our CFC measurement uses three identical narrow band filters with center frequencies located on n-m, n, and n+m. The method therefore is free from the bandwidth dilemma.

**Comparison with Existing Method(s):** The method was tested on diagnostic artificial signals modeled on local field potentials and compared with four established PAC detection algorithms. While the proposed method detected the simulated PAC in high frequency resolution, the other methods detected with poor frequency resolution, or completely missed the PAC.

**Conclusion:** Using the proposed triplet-filter banks instead of wideband filtering allows for high resolution detection of PAC. Moreover, the method successfully detected PAC in wide range of modulation frequency. Finally, bandwidth is not chosen subjectively in our new method which makes the comparison of PAC more convenient among different studies.

230 words

**Funding:** This research is founded by an Odysseus grant from the Flemish organizations for science (FWO) awarded to Cees van Leeuwen.

**Introduction**

Brain activity generates an electric field, of which the oscillatory components extend over several frequency bands, ranging from slow: Delta (1-4Hz), Theta (4-8Hz), and Alpha (8-13Hz), to fast: Beta (13-30Hz) and Gamma (>30Hz). Oscillatory activity is generated by populations of interconnected neurons (Reiner and Anderson, 1990; Buzsáki and Draguhn, 2004), is understood to regulate information flow between brain regions (Varela et al., 2001; Schnitzler and Gross, 2005; Fries, 2015), and to play a role in sensory and motor control functions, cognitive processes such as attention and memory, and emotional states (Başar et al., 2001; Cantero and Atienza, 2005).

Oscillatory activity bands interact with each other. In particular, the phase of the slow oscillations is known to modulate the amplitude of the fast ones; a phenomenon known as Phase-Amplitude Coupling (PAC). For instance, theta band phase modulates the amplitude of Gamma band activity in human EEG and rodents LFP (Demiralp et al., 2007; Newman et al., 2013). PAC has been related to a range of human information processing functions and mental states (Canolty et al., 2006; Schutter and Knyazev, 2012), e.g. increasing working memory load correlates with PAC (Sauseng et al, 2004).

Several methods have been developed for assessing PAC (for reviews, see (Tort et al., 2010a)), each with their own advantages and limitations (Berman et al., 2012; Aru et al., 2015, Hyafil, 2015). Typically, components are registered in the time-frequency domain based on band-pass filtering, using, for example, the Morlet wavelet transform. This, however, leads to a dilemma with respect to the choice of filter bandwidth. Consider the modulation of an oscillatory component at frequency $n$ by an oscillation at frequency $m$. Their modulatory components exist at frequencies $n-m$ and $n+m$. While narrow-band filtering around $n$ is optimal for detecting the frequency component, this misses the modulatory components. To detect amplitude modulation around the center frequency $n$, a minimum bandwidth of $2m$ is needed. However, such wide band filtering leads to the inclusion of noise and thus a drop in the accuracy of detection (Aru et al., 2015). While still suitable for detecting PAC measures when $m$ remains in the Delta and Theta range, this choice of bandwidth will reduce drastically the accuracy of detecting PAC when $m$ is in the Alpha and/or Beta frequency bands.

We introduce *Modulatory Component Analysis* (MCA) as a new approach to PAC detection, based on using a combination of narrow band filters centered at the frequency of oscillations and their modulations. The narrow band filters improve signal to noise ratio (SNR) through reducing the noise. MCA therefore is free from the bandwidth dilemma. We will compare MCA with four other PAC detection methods, namely Envelope Phase Synchronization (EPS), Mean Vector Length (MVL), Coherence Value (CV) and Kullback-Leibler Distance (KLD).

**Methods**

*Definition of signals and operator*

The signals and the operator used in the MCA algorithm are as follows:

1. The signal $Z_X(t)$ is the analytic representation of the real valued signal X(t), defined as:

$$Z_X(t) = \hat{X}(t) + iX(t) \qquad (1)$$

   Where $\hat{X}(t) = Hilbert(X(t))$. The analytic representation of a real valued signal yields a complex signal (i.e. $Z_X(t) = a_X(t)e^{i\varphi_X(t)}$) which contains instantaneous amplitude and instantaneous phase information.

2. The signal $a_X(t)$ is the instantaneous amplitude of signal $X(t)$ and is obtained by extracting the amplitude information of its analytical representation:

$$a_X(t) = |Z_X(t)| \qquad (2)$$

3. The signal $\varphi_X(t)$ is the instantaneous phase of signal $X(t)$, obtained by extracting the phase information of the analytical representation of signal $X(t)$:

$$\varphi_X(t) = \angle Z_X(t) \qquad (3)$$

4. The signal $f_X(t)$, the instantaneous frequency of signal $X(t)$, is the time derivative of its instantaneous phase, defined as:

$$f_X(t) = \frac{1}{2\pi} \frac{d\angle Z_X(t)}{dt} \qquad (4)$$

5. The signal $X_n(t)$ is the real valued Gabor filtered signal, i.e. a band-pass filter at center frequency *n*, defined as:

$$X_n(t) = e^{-(\kappa.n.t)^2} \cos(2\pi n t) * X(t) \qquad (5)$$

Where "*" is the convolution operator and "κ" is the bandwidth parameter. In this study κ=1 which results in an approximate bandwidth of 1Hz. Note that Gabor filters are used in Morlet wavelet analysis as well.

6. The signal $X_{m,n}(t)$ is the "n" Hz frequency component of signal $X(t)$, together with the assumed modulatory information located at "*m+n*" and "*m-n*" Hz frequency. The signal $X_{m,n}(t)$ defined as:

$$X_{m,n}(t) = X_{n-m}(t) + 2X_n(t) + X_{m+n}(t) \qquad (6)$$

$X_{m,n}(t)$ is the most critical part of the MCA algorithm: Applying the analytical signal operator on $X_{m,n}(t)$ results in high resolution instantaneous amplitude and instantaneous frequency information at *m* Hz. The reason is that the *m* Hz modulatory information around the carrier frequency of *n* Hz is perfectly preserved while noise resulting from using a large bandwidth is excluded. This results in high SNR and accordingly high accuracy in the extraction of the modulatory information.

7. The Phase Locking Value (PLV) between two time series $U(t)$ and $V(t)$ is calculated as(Lachaux et al., 1999):

$$PLV(U(t), V(t)) = \left| < e^{i(\varphi_U(t) - \varphi_V(t))} > \right| \qquad (7)$$

The *PLV* index varies between zero and one, where the value one indicates perfect phase synchronization between the time series $U(t)$ and $V(t)$ and zero value indicates absence of phase synchronization between them.

*MCA Algorithm for PAC*

A schematic of the MCA algorithm is given in Figure 1. The algorithm extracts the instantaneous amplitude, $a_{X_{m,n}}(t)$, from narrow band signals, $X_{m,n}(t)$, as defined in Equations (2) and (6). The slower activity, $X_m(t)$, is obtained by Gabor filtering of the raw signal, $X(t)$, at center frequency *m*. The MCA algorithm for measuring the PAC between the fast oscillatory activity at frequency *n* and the slower oscillatory activity at frequency *m* is expressed as follows:

$$MCA - PAC_{m,n}(X(t)) = PLV(X_m(t), a_{X_{m,n}}(t)) \qquad (8)$$

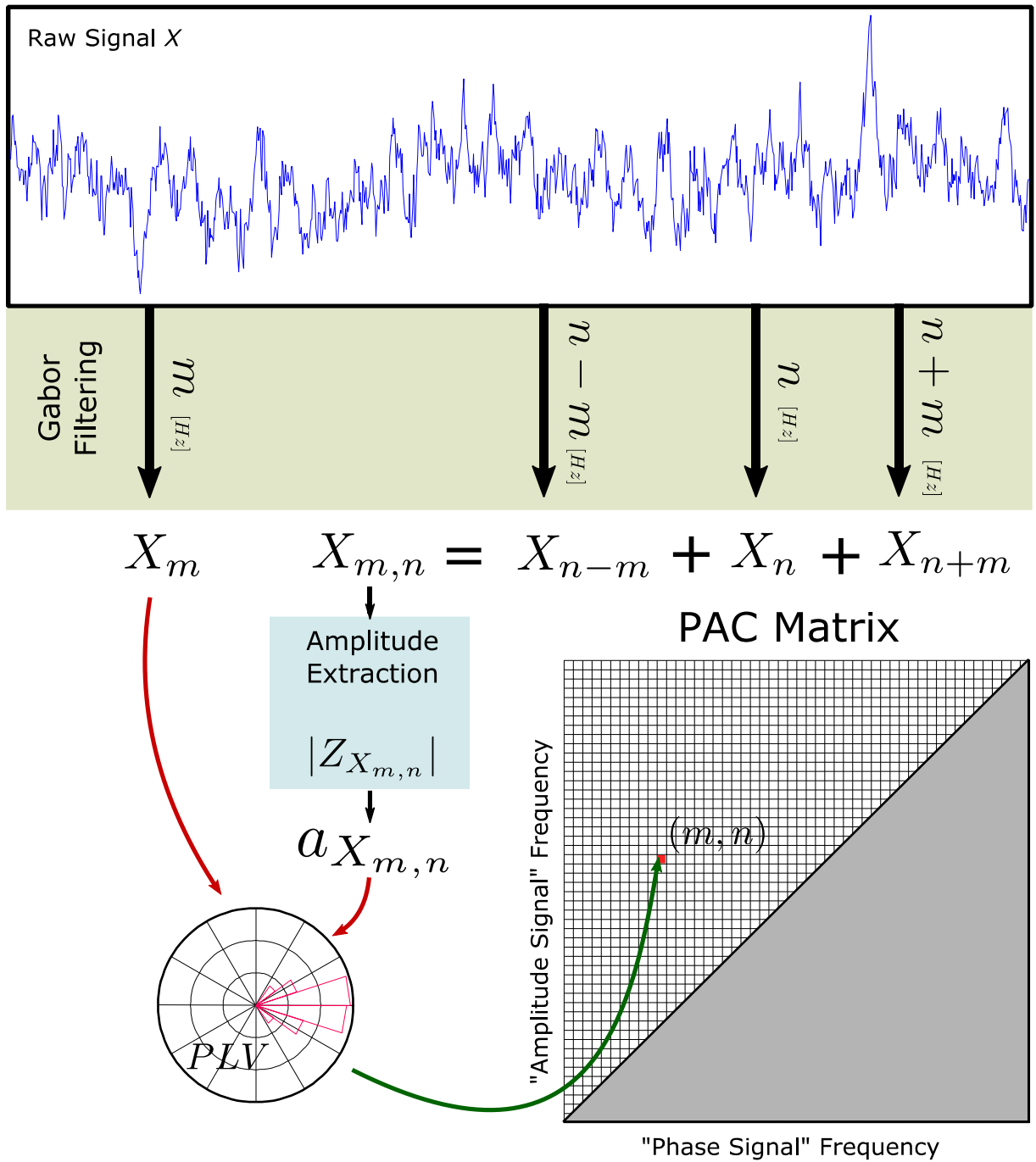

*Figure 1 - Illustration of the Modulatory Component Algorithm (MCA)* MCA-PAC offers a Phase-Amplitude Coupling measure. Signals $X_m(t)$ and $X_{m,n}(t)$ are obtained though illustrated Gabor filtering steps. In MCA-PAC the synchrony between the slow activity $X_m(t)$ and the instantaneous amplitude of the fast activity ($a_{X_{m,n}}(t)$) is evaluated using the PLV function.

*Review of Four Commonly Used PAC Methods*

Here, we briefly review four established measures used for detecting PAC: envelope phase synchronization (Cohen et al. 2008), mean vector length (Canolty et al. 2006), coherence value (Colgin et al., 2009), and Kullback-Leibler distance (Tort et al., 2010). The formulations of the algorithms are adapted to the notations introduced in the method section.

a) *Envelope Phase Synchronization (EPS)*

To measure PAC between the *m* and *n* frequencies, Cohen et al. (2008) calculated the PLV (Equation (7)) between the slow oscillation and the instantaneous amplitude of the fast oscillation. The slow and fast oscillations, $X_m(t)$ and $X_n(t)$ respectively, are obtained using the Gabor filters (Equation (5)). Cohen et. al. (2008) implemented the Gabor filters using the Morlet Wavelet. The instantaneous amplitude of the fast oscillation, $a_{X_n}(t)$, is calcualted through the analytic tarsformation (Equation (2)). Eventually, the EPS algorithm calculates PAC as follows:

$$EPS_{m,n}(X(t)) = PLV(X_m(t), a_{X_n}(t)) \qquad (9)$$

b) *Mean Vector Length (MVL)*

Canolty et al. (2006) originally introduced the "Mean Vector Length" method to measure PAC. The authors defined a time variable vector in the complex plane as $a_{X_n}(t)e^{i\varphi_{X_m}(t)}$, in which $a_{X_n}(t)$ is the instantaneous amplitude of the fast oscillation at frequency *n*, and $\varphi_{X_m}(t)$ is the instantaneous phase of the slow oscillation at frequency *m*. In the absence of the PAC, these vectors have a uniform circular density and are symmetric around zero. This leads to small values for the mean of the vector lengths. Any systemic relation between the phase and amplitude will increase the density of the vector points for a specific phase, resulting in an increase of the mean vector length. The measure can be described as follows:

$$MVL_{m,n}(X(t)) = |< a_{X_n}(t)e^{i\varphi_{X_m}(t)} >_t| \qquad (10)$$

In which the signals $X_m(t)$, $X_n(t)$, $a_{X_n}(t)$ and $\varphi_{X_m}(t)$ are calculated according to Equations (5), (5),(2) and (3) respectively.

c) *The Coherence Value (CV)*

Colgin et al., (2009) calculated the frequency coherence between the unfiltered signal, $X(t)$, and the instantaneous amplitude of the fast oscillation. The fast oscillation is obtained using a Gabor filter centered at frequency $n$. According to this measure, the coherence value at frequency $m$ gives the PAC strength between the $m$ and $n$ frequency components. The measure assesses the PAC strength as follows:

$$CV_{m,n}(X(t)) = Coh_m\left(X(t), a_{X_n}(t)\right) \qquad (11)$$

In which the instantaneous amplitude $a_{X_n}(t)$ is calculated according to Equation (2).

d) *Kullback-Leibler Distance (KLD)*

To measure PAC, Tort et al. (2010) evaluated the statistical relation between phase and amplitude time series. Initially, the instantaneous phase of the slow oscillation ($\varphi_{X_m}(t)$) and the instantaneous amplitude of the fast oscillation ($a_{X_n}(t)$), are extracted. Next, the discrete probability distribution of $a_{X_n}(t)$ among different values of $\varphi_{X_m}(t)$ is calculated. Tort et al. (2010), binned the phase to reduce the number of phase points. Binning was also applied to calculate this measure in the current study (the number of bins was 50). When the obtained distribution is uniform, this implies that there is no systematic relation between the phase and amplitude, i.e. absence of PAC. The amount of deviation from the uniform distribution gives the PAC strength. The Kullback–Leibler Distance (KLD) function measures the deviation as follows:

$$KLD_{m,n}(X(t)) = 1 - \frac{H_N(P_{a_{X_n}}(\varphi_{X_m}))}{\log(N)} \qquad (12)$$

in which $N$ is the number of phase bins and $H_N(P)$ is the entropy function, defined as:

$$H_N(P) = \sum_{l=1}^{N} P(l)\log(P(l)) \qquad (13)$$

In Equation (13, $P_{a_{X_n}}(\varphi_{X_m})$ is the probability distribution of fast activity instantaneous amplitude at frequency $n$, for binned phase values of slow activity at frequency $m$.

For the sake of comparison with the MCA method, the four algorithms (*a-d*) were all coded in MATLAB. The Morlet wavelet width-parameter was set to 4 for all four algorithms (Torrence and Compo, 1998). This value gives the highest allowed bandwidth for each center frequency.

*Pure-PAC Signals*

Tests were conducted using an artificial LFP-like signal generated in MATLAB. A set of pure PAC signals were generated as following:

$$X_{PAC}(t) = \sin(2\pi mt) + (0.5 + AMI \times \sin(2\pi mt)) \times \cos(2\pi nt) + PinkNoise(t) \qquad (14)$$

where *AMI* is the amplitude modulation index that determines amplitude coupling strength; *m* and *n* are the frequencies of the modulating and modulated oscillations, respectively.

In the set, four pairs of coupling frequencies are considered: $(m,n)\epsilon\{(8,45), (12,45), (20,45)\ and\ (30,45)\}$. These were chosen based on empirical studies in LFP, which reported modulations between Alpha and Gamma (Chen et al., 2015), and Beta and Gamma bands (Dejean et al., 2011; Lu et al., 2015; Nair et al., 2016). The time length of the resulting six simulated signal is 10 seconds, discretized by time steps of 0.001 seconds. Pink noise ($1/f$) was added to each of the simulated signals, reflecting the typical power spectrum of LFP (Pritchard, 1992; Bédard and Destexhe, 2009; Dehghani et al., 2010).

*AMI* parameters was manually adjusted to render the power spectrum density (PSD) of the target signals to be similar to PSD of LFP recordings. Using *AMI* = 0.25 and power of pink noise = 6250 Watt, PSD of $X_{PAC}(t)$ is computed using the Welch method. A sliding Hanning time window of length equal to 16348 data point was applied to the signals, allowing 25% window overlap. The power of individual signals was 630±10 Watt. SNR of each simulated signal was around 0.1 which is comparable to measured values of SNR in LFP (Łęski et al., 2013; Vinck et al., 2015).

Figure 2 shows the PSD of the PAC signals used in our simulations. Peaks appear at the chosen frequency locations, {8*Hz*, 45*Hz*}, {12*Hz*, 45*Hz*}, {20*Hz*, 45*Hz*}, and {30*Hz*, 45*Hz*}, which corresponds to prominent oscillatory components at Alpha and Gamma (Figure 7a and Figure 7b) (Chen et al., 2015) and Beta and Gamma bands (Figure 7c and Figure 7d) (Dejean et al., 2011; Lu et al., 2015; Nair et al., 2016) in LFP.

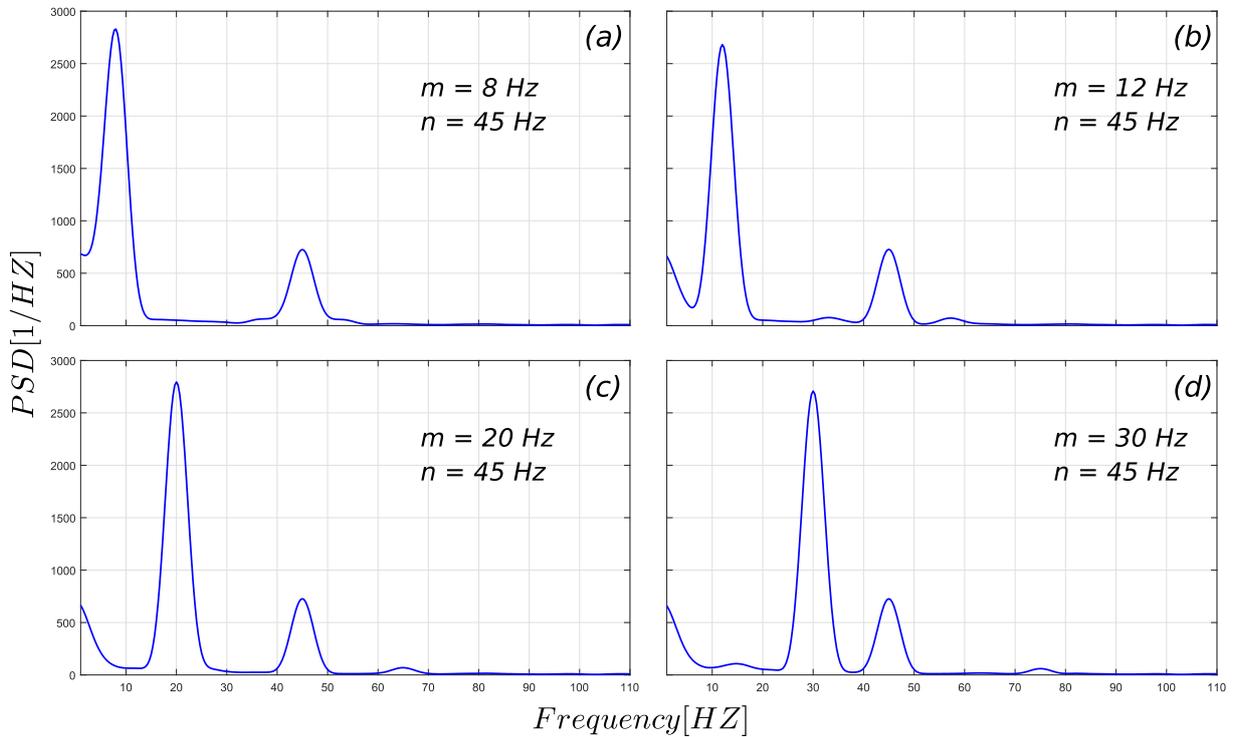

*Figure 2 – **The Power Spectral Density of the artificial LFP-like signals.** a) The PSD of PAC Signal for the (8,45) frequency pair (b) The PSD of PAC signal for the (12,45) frequency pair (c) The PSD of PAC signals for the (20,45) frequency pair (d) The PSD of PAC signal for the (30,45) frequency pair.*

*Calculation of PAC Matrices*

Each PAC measure calculates the coupling strength as a real positive number for all frequency combinations of *(m,n)* where $m, n \in \{1,2,3,...,50\} Hz$. However, PAC is only meaningful when *m < n*. Therefore, the corresponding coupling value is manually set to zero when *n ≥ m*. As a result, each PAC measure gives a [50x50] triangular matrix in which the columns and rows represent modulatory (*m*) and modulated (*n*) components, respectively. The PAC matrices are calculated for all the signals in the frequency pairs set which gives four PAC matrices per method. For comparison, PAC values in each matric are normalized by dividing them by the largest element of that matric.

# Results

*MCA algorithm*

Results of the MCA algorithm are presented in Figure 3, in which the PAC matrices for the signals in the frequency pairs set are shown in Panels a-d. The global maximum (white boxes in Figure 3) indicates the modulated and modulating frequencies. Thus, the algorithm adequately detects the PAC. Sporadic high values (although lower than the global maximum) appear elsewhere, e.g., at (37, 45) *Hz* in Panel a. These occur at intersections with (higher) harmonics of the modulating and modulating frequencies, e.g. 37 = 45-8*Hz*. The results, therefore, show that MCA clearly detects PAC at high resolution, for the wide range of modulating frequency *m*.

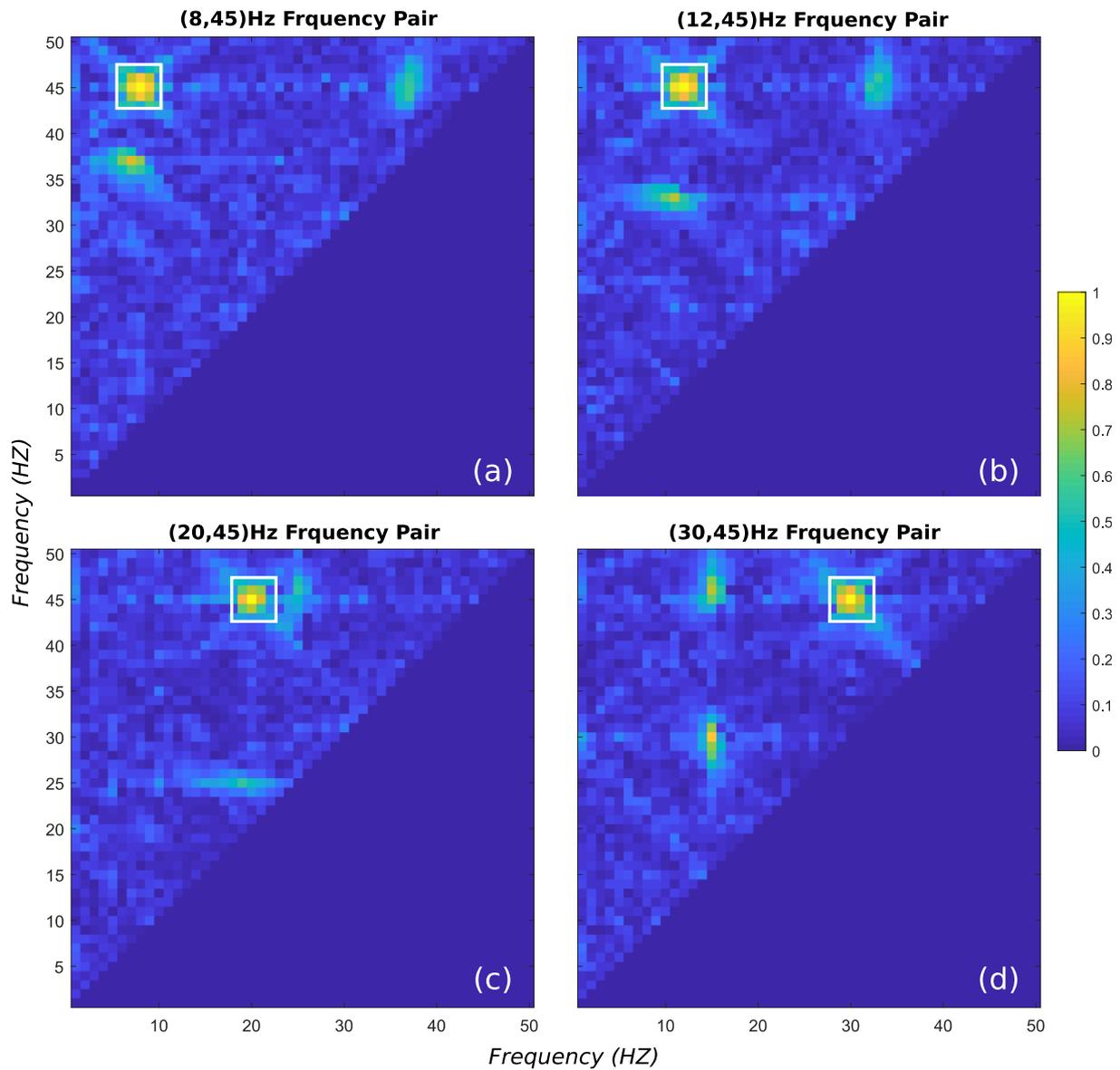

*Figure 3 – **Performance of the MCA-PAC measure on the artificial LFP-like signals.** Panels a to d show PAC detection matrices of PAC signals computed for four coupling frequency pairs. The color bar indicates the intensity of the PAC for which blue indicates the lowest and yellow indicates the highest level of coupling.*

In what follows, we will compare the performance of MCA to that of Envelope Phase Synchronization (EPS) and three other methods offer PAC measures. Each measure was applied to the artificial PAC signals to investigate their liability for precise PAC detection.

*Envelope Phase Synchronization (EPS)*

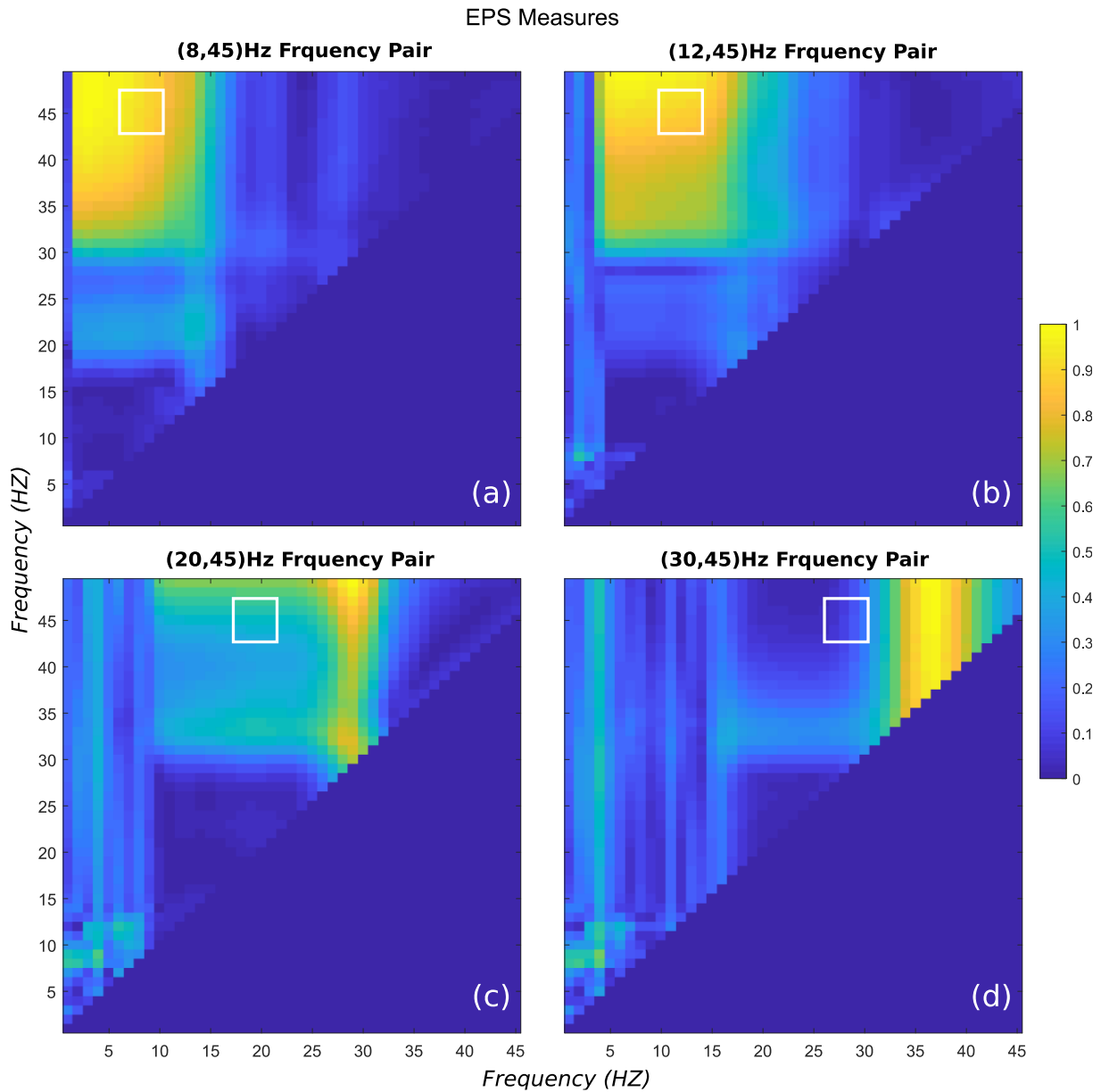

*Figure 4 - Performance of the Envelope Phase Synchronization (EPS) measure on the artificial LFP-like signals. Panels a to d show PAC detection matrices of PAC signals computed for four coupling frequency pairs. The color bar indicates the intensity of the PAC for which blue indicates the lowest and yellow indicates the highest level of coupling.*

Figure 4 shows PAC matrices calculated by EPS in which panels a-d show the results for PAC signals in frequency pairs set. The highest PAC was reported around the correct location for the (8,45) frequency pair (Panel a). The detection of PAC at frequency pair (12,45) is not accurate, but still acceptable (Panel b). However, the EPS measure performs poorly for the (20,45) and (30,45) pairs (Panels c and d). Thus, PAC detectability is poor in particular when the modulating activity is fast. The method performed worse than the MCA-PAC. The difference is due to the Morlet wavelet filters which is the only difference between the two methods. The filter passed wide-band signals including extraneous ones. The wide-band characteristics blurred the EPS-PAC matrices and impaired the detectability where m is in the higher frequency band.

*Mean Vector Length (MVL)*

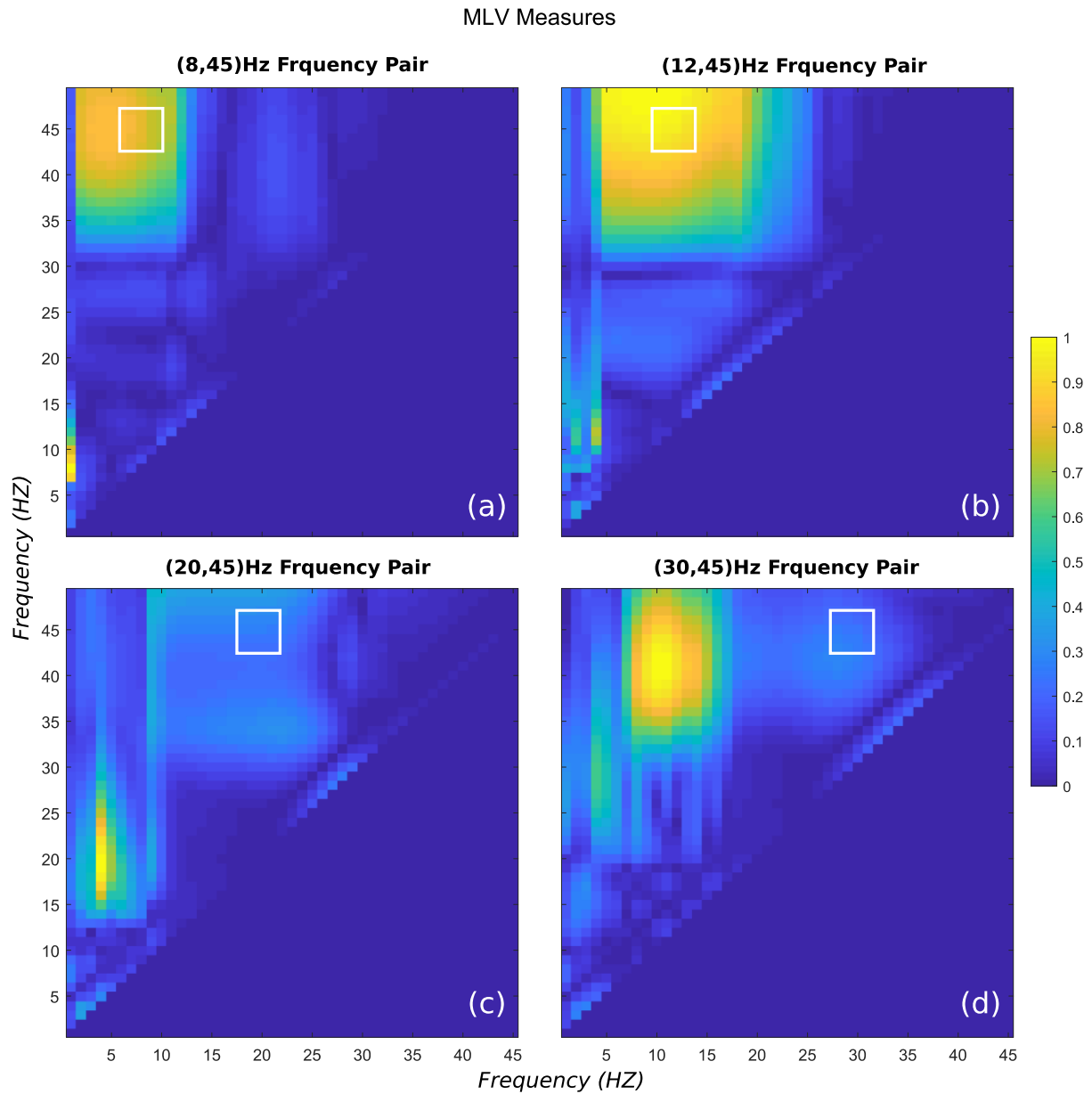

*Figure 5 – Performance of the Mean Vector Length (MVL) measure on the artificial LFP-like signals. Panels a to d show PAC detection matrices of PAC signals computed for four coupling frequency pairs. The color bar indicates the intensity of the PAC for which blue indicates the lowest and yellow indicates the highest level of coupling.*

Figure 5 shows results for the standardized Mean Vector Length (MVL) measure (again, PAC signals in frequency pairs set in Panels a-d). For the (8, 45) and (12, 45) pairs, PACs were detected with acceptable level of accuracy (Panels a and b), however, this was not the case for other two pairs (Panels c and d). The performance of MVL in detecting PAC is comparable to that of EPS, and worse than MCA-PAC. The difference is also due to the Morlet wavelet filter.

*Coherence Value(CV)*

Figure 6 shows the performance of Coherence Value (CV) measure. The CV reported PAC for the (8, 45) and (12,45) pair at the approximately the correct location (Panel a and b). Although the accuracy for these two frequency pairs is poor. For the (20, 45) and (30, 45) pair, CV failed to perform correctly; A false PAC global maximum occurred elsewhere (Panels c and d). These somewhat irregular results might occur because CV is calculated between band passed fast activity and an unfiltered signal (Cf. Equation (12)). Whereas this method is faster than the other methods, it is considerably less reliable than the others, in particular when compared to MCA.

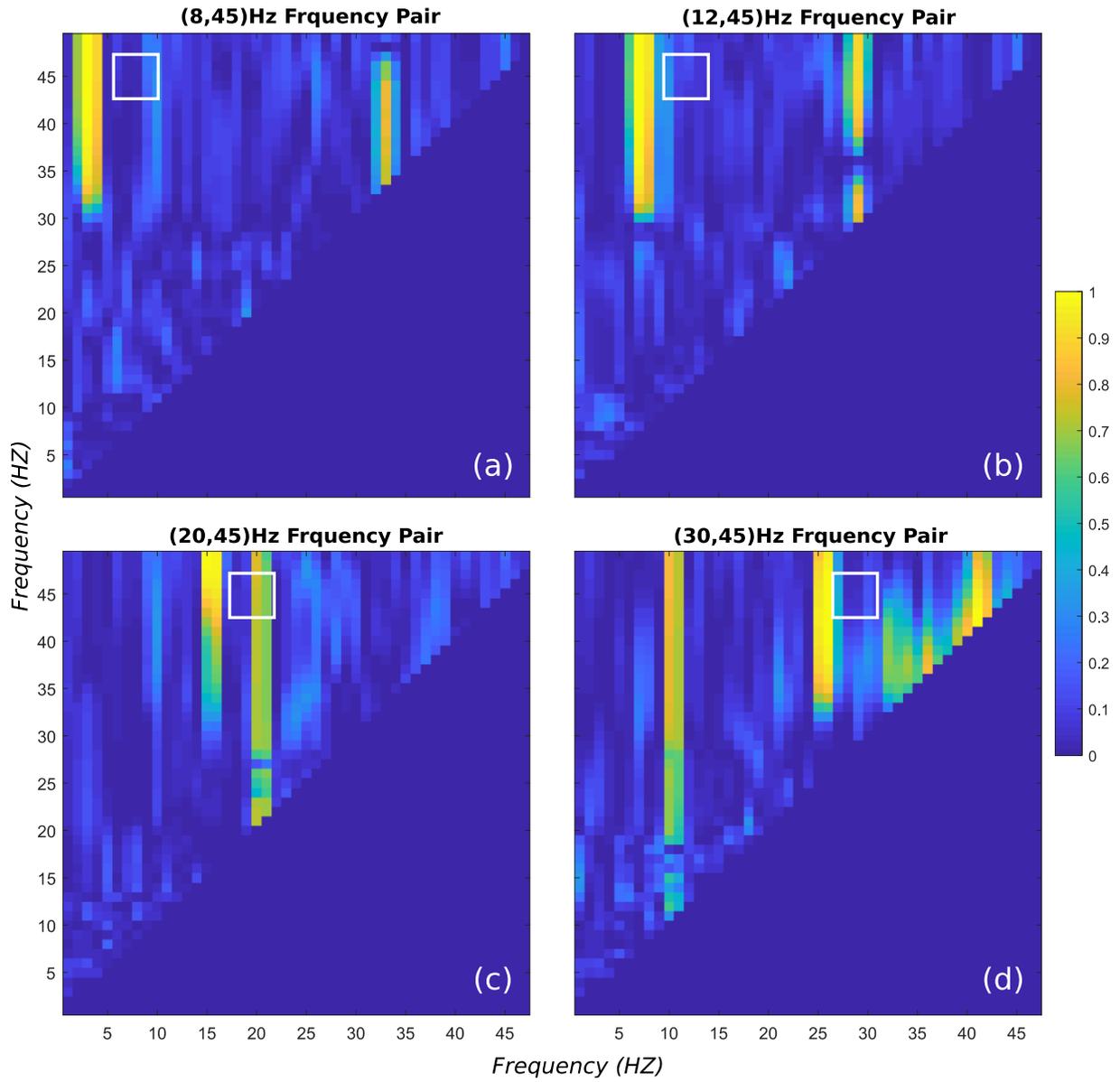

*Figure 6 - **Performance of the Coherence Value (CV) measure on the Artificial LFP-Like signals.** Panels a to d show PAC detection matrices of PAC signals computed for four coupling frequency pairs. The color bar indicates the intensity of the PAC for which blue indicates the lowest and yellow indicates the highest level of coupling.*

*Kullback-Leibler Distance (KLD)*

KLD Measures

(a) (8,45)Hz Frquency Pair
(b) (12,45)Hz Frquency Pair
(c) (20,45)Hz Frquency Pair
(d) (30,45)Hz Frquency Pair

*Frequency (HZ)*

*Figure 7 - **Performance of the Kullback-Leibler Distance (KLD) measure on the Artificial LFP-Like signals** Panels a to d show PAC detection matrices of PAC signals computed for four coupling frequency pairs. The color bar indicates the intensity of the PAC for which blue indicates the lowest and yellow indicates the highest level of coupling.*

Figure 7 presents the results for the Kullback-Leibler Distance (KLD) measure. PAC is approximately correctly detected for the (8,45) and (12,45) pairs (Panels a and b); it reports false PAC in the fast frequency pairs (Panels c and d). The performance of this method is comparable with EPS and MVL methods.

**Discussion**

For detecting Phase Amplitude Coupling (PAC), we introduced Modulatory Component Analysis (MCA), a method based on clusters of narrow band filters. MCA accurately detected PAC embedded in artificial LFP signals in high frequency resolution. In contrast, four established PAC measures detected PAC only in low resolution when the modulating frequency was low and failed completely when the modulating frequency was high.

The source of the problems in the established methods is the bandwidth of the Morlet wavelet filter. In order to balance time and frequency resolution under conditions of uncertainty, the bandwidth, specified by the "wavelet width" parameter, varies proportionally to the center frequency. Higher bandwidths are utilized for higher center frequencies, in order to secure accuracy in temporal resolution, but this leads to lower frequency resolution. Thus, for high modulating frequencies which require larger bandwidths, the modulation information is more likely to be missed.

Because of this, the scope of traditional PAC analysis in Wavelet-based algorithms is restricted. For instance, to detect PAC between 20 and 45$Hz$, the bandwidth should be at least 40$Hz$. The half-gain bandwidth of the Morlet wavelet filter at the central frequency 45$Hz$ is 30$Hz$. The filter, therefore, is too narrow to cover the requisite range. Application of a wide(r)-band filter, as suggested by (Berman et al., 2012), would include the modulatory information, yet this would introduce another problem: the filter will also pass extraneous oscillatory components. These will cause perturbations passing through the filter, thereby reducing the accuracy of modulation detection. MCA eliminates this dilemma by selecting modulatory information, such that with modulatory frequency m and modulated frequency n, the frequencies n-m, n, and n+m are narrowly passed.

The application of a cluster of narrow-band filters as advocated by our approach, improves detection of PAC, as clearly shown in the comparison between MCA-PAC and EPS. The MCA algorithm, by detaching the link between center frequency and bandwidth enlarges the scope of detectable PAC to a broader range of frequency pairs. Clustered filtering is by no means privy to our approach. The performance of MVL, CV, and KLD methods could likewise be improved using a filter cluster.

MCA algorithm can facilitate the comparison among different studies. In current studies, bandwidth variable is chosen subjectively, mainly based on the frequency bands under the study. For instance, the bandwidth variable chosen for Theta-Gamma PAC investigation is different from the same type of research for which Delta-Gamma coupling is studied. This will introduce unintended bias into the intensity of the couplings, even if both used datasets set are identical. MCA algorithm the bandwidth variable is inessential and therefore the comparison becomes fairly straightforward.

The narrow band filtering works the best for signals with distinctive frequency peaks. In real data, the oscillatory components will have broader spectral peaks than those in the simulated data. This may compromise the high frequency resolution of MCA algorithm, e.g., from 1Hz to 4Hz. Nevertheless, the resolution is much finer than that of any method involving the Morlet wavelet. As a result, shifts in coupling frequency pairs, for instance during cognitive or visual tasks (Voytek et al., 2010), could be tracked with MCA. This gives MCA a definite advantage over wavelet-based analysis methods for observing cross-frequency coupling.

A downside of the MCA algorithm is its computational cost. The number of filtering stages it requires is in the order of $N^2/2$, where N is the size of the CFC matrix. For the other methods reviewed, the number of filtering steps is of order N; this makes the MCA algorithm $N/2$ times more computationally expensive. The cost could be reduced by means of parallelization of computation. The filtering steps in MCA are independent, and thus ideally suitable for parallelization. For example, after parallelization in Matlab with 12 processors the method is only 3 or 4 times more expensive compared to current PAC measures.